# Dark Matter Searches at the Large Hadron Collider

S. Y. Hoh[1, a)], J. R. Komaragiri[1, b)] and W. A. T. Wan Abdullah[1, c)]

[1]National Centre for Particle Physics (NCPP), Level 4, Research Management and, Innovation Complex, University of Malaya, 50603 Kuala Lumpur, Malaysia.

a)Corresponding author: siew.yan.hoh@cern.ch
b)jyothsna@um.edu.my
c)wat@um.edu.my

**Abstract.** Dark Matter is a hypothetical particle proposed to explain the missing matter expected from the cosmological observation. The motivation of Dark Matter is overwhelming however as it is mainly deduced from its gravitational interaction, for it does little to pinpoint what Dark Matter really is. In WIMPs Miracle, weakly interactive massive particle being the Dark Matter candidate is correctly producing the current thermal relic density at weak scale, implying the possibility of producing and detecting it in Large Hadron Collider. Assuming WIMPs being the maverick particle within collider, it is expected to be pair produced in association with a Standard Model particle. The presence of the WIMPs pair is inferred from the Missing Transverse Energy (MET) which is the vector sum of the imbalance in the transverse momentum plane recoils a Standard Model Particle. The collider is able to produce light mass Dark Matter which the traditional detection fail to detect due to the small momentum transfer involved in the interaction; on the other hand, the traditional detection is robust in detecting a higher Dark matter masses but the collider is suffered from the parton distribution function suppression. Topologically the processes are similar to the scattering processes in the direct detection thus complementary to the traditional Dark Matter detection. The collider searches are strongly motivated as the results are usually translated to the annihilation and scattering rates at more traditional Dark Matter-oriented experiments, thus a concordance approach is adapted. An overview of Dark Matter searches at the Large Hadron Collider will be covered in this paper.

## INTRODUCTION

The world renowned particle collider known as the Large Hadron Collider (LHC) at CERN, Geneva at Switzerland has claimed triumph over the discovery of Higgs Boson in 2012 confirming the rigidity of Standard Model. However, particularly based on the cosmological study [1], ~5% of the universe is surprisingly made up of Standard Model particle(SM), and the rest made up of unknown entities, ~26.4% of Dark Matter (DM) and ~68.5% of Dark Energy. Since LHC is able to produce pre-universe environment, it is highly expected that DM will be produced within the large hermicity of the detector thus propelling the discovery of DM.

     DM begin to gain recognition in 1933 inspired by Frizt Zwicky who studied the relative velocity of each cluster member in the Coma Cluster [2]. Armed with Virial Theorem he predicted the total mass of the Cluster, which to his surprise the mass inferred from the relative velocity is 400 times the mass of the visible star in galaxies in the cluster. The observation was soon confirmed by a similar measurement of the Virgo Cluster carried out by Smith [3]. A consistent contradiction was further substantiated by Vera Rubin who found that most of the galaxy rotational velocity remained constant at larger distance, contradicting the Keplerian prediction [4]. Despite there were theoretical efforts carried out in spirit to explain the observation by modifying the Newtonian mechanics rather than introducing a new form of matter [5], inevitably a new form of particle is strongly motivated due to the compelling evidence of IE0657-558 [6]. Evidently the non-baryonic matter does not interact with the baryonic matter. From the total density of mass distribution measured it showed the total mass moved ballistically after the collision, indicating DM self-interactions were weak in nature.

## WIMP AS DARK MATTER CANDIDATE

A wide range of DM candidates such as MACHO (Massive Compact Halo Object) and Primordial Black Hole were excluded based on the cosmic microwave background and the large-structure formation study which subsequently theorized ΛCDM (Cold Dark Matter) [7]. The DM particle is known to be massive due to its non-relativistic speed, invisible, electromagnetically neutral, and only interact with weak interaction and gravity. Initially the neutrino was suspected to be DM candidate but it was immediately dismissed as it was not massive enough. Consequently, the Standard Model does not provide a viable DM candidate. However in physics beyond Standard Model such particle happens to be naturally motivated by a wide spectrum of models which attempt to solve gravity and strong CP problem. Besides sterile neutrinos and Axions, the compelling DM candidate is WIMP (Weakly Interacting Massive Particle).

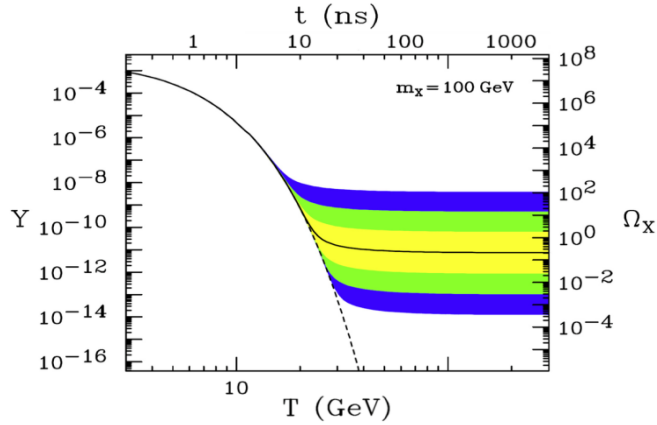

**FIGURE 1**: The number density Y (left) of a 100GeV, P-wave annihilating Dark Matter particle evolves as a function of time t (top) and temperature T (bottom). The solid contour is the annihilation cross section yields the correct relic density Ωχ (right), which is~0.23 [8].

The early universe was full of radiation fuelled by the constant pair-production and annihilation of particles while establishing a thermal equilibrium. The WIMP was assumed to produce and annihilate in the same fashion as other particles until it decouples from the thermal equilibrium as the annihilation process diminishes due to the subsequent universe expansion. Figure 1 shows the evolution of thermal relic density for WIMP with mass $M_\chi$ =100 GeV, when the universe was expanded and cooled at T < $M_\chi$, the DM was "frozen-out" and left over, manifestly the comoving number density Y remained constant and produced the present thermal relic density $\Omega_\chi$. By solving the Boltzmann equation the relic density implied that the average DM annihilation cross section was ~< σAv > = 1pb which turns out to be the typical cross section the LHC is currently producing. Furthermore a simple dimensional analysis suggests if the correct thermal relic was made up of DM, the DM particle falls on the weak-scale mass range of 100 GeV - 1 TeV. This connection between Cosmology and Particle Physics therefore has established the WIMP Miracle. Naturally the WIMP Miracle implied many model providing viable DM candidates. For instance, the DM candidates are the lightest neutralino from weak-scale Supersymmetry and Kaluza-Klein photon from extra dimensions [8].

## MODEL INDEPENDENT COLLIDER SEARCHES

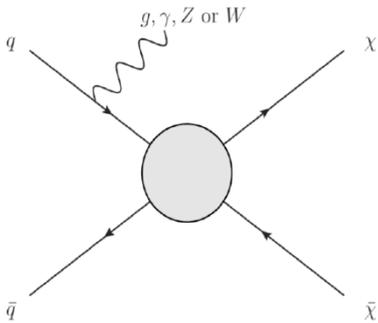

| Name | Initial state | Type | Operator |
|------|---------------|------|----------|
| D1 | $qq$ | scalar | $\frac{m_q}{M_*^3}\bar{\chi}\chi\bar{q}q$ |
| D5 | $qq$ | vector | $\frac{1}{M_*^2}\bar{\chi}\gamma^\mu\chi\bar{q}\gamma_\mu q$ |
| D8 | $qq$ | axial-vector | $\frac{1}{M_*^2}\bar{\chi}\gamma^\mu\gamma^5\chi\bar{q}\gamma_\mu\gamma^\mu q$ |
| D9 | $qq$ | tensor | $\frac{1}{M_*^2}\bar{\chi}\sigma^{\mu\nu}\chi\bar{q}\sigma_{\mu\nu}q$ |
| D11 | $gg$ | scalar | $\frac{1}{4M_*^3}\bar{\chi}\chi\alpha_s(G^s_{\mu\nu})^2$ |

**FIGURE 2**: In EFT language, the vertex between DM and SM interaction is covered with a blob indicating M∗ (left) and a list of potential DM operators assuming DM is Dirac fermion (right) [9].

In accordance to the thermal relic freeze out framework, the DM has to minimally interact weakly with SM in order to produce the current thermal relic. The interaction of DM with SM is heavily model-dependent because the coupling usually is restricted by gauge invariance and other symmetries especially for a higher spin DM particle. It may be useful to assume a more generic interaction rather than model-oriented in hope of understanding which is truly generic to Dark Matter physics. In the model-independent approach, with the assumption that DM exists and stable, the interactions between DM and SM are parametrized by a set of effective non-renormalizable operators which mimic the effect of heavy mediator. The factorization of heavy mediator generates a set of DM contact operator shown in Figure 2 which effectively describe the interaction [10]. The Effective Field Theory (EFT) approach is justified whenever there is a clear separation between the energy scale M∗ and the underlying microscopic interaction of the process. For example, for indirect DM searches the annihilation of non-relativistic DM particles occurs with momentum transfer $Q^2$ of the order of $M\chi$; in direct searches, the $Q^2$ is of few orders of tens of keV in the scattering process of DM on heavy nuclei. Therefore it is possible to carry out an effective description in terms of DM operator with a Ultra-Violet (UV) cutoff larger than the typical $Q^2$ to limit on $M_*$.

However for LHC searches it is dramatically different from the other DM searches. The $Q^2$ involved can be very high that the EFT description is no longer valid. Nevertheless under some condition the EFT description still can be valid if the energy scale involving the DM and SM process is smaller compared to the energy scale of heavy mediator [10]. A light mediator version or a simplified model where it assumes only one new particle with mass $M_{med}$ which may come from the dark sector accurately describes the DM and SM interaction. After all EFT is an approximation of simplified model corresponding to expanding the propagator of the heavy mediator in powers of $Q^2_{tr}/M^2_{med}$ truncating at lowest order.

Since most of the detection technique from direct, indirect and LHC required an interaction of WIMP with the SM, such interaction may be generated by the same operator. Therefore from the Particle Physics point of view, the collider searches served as complementary searches to the traditional DM searches by appropriately correcting the kinematics [9] according to the interaction channels. Figure 3 summarizes all the DM detection technique exploited by various DM searches.

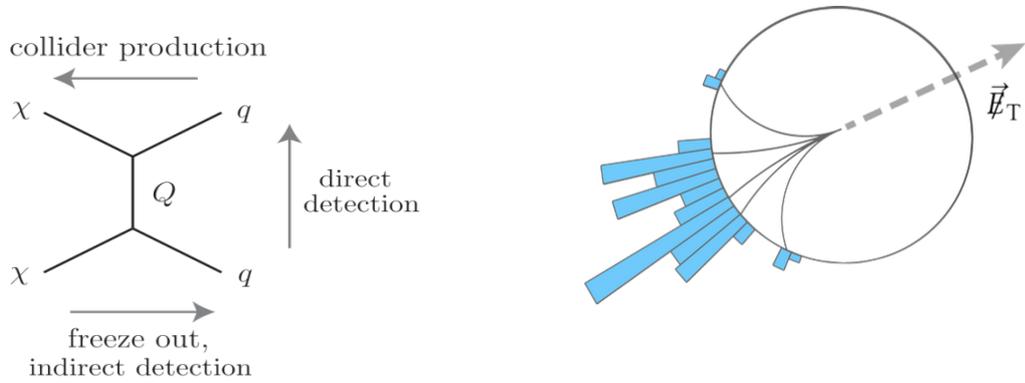

**FIGURE 3**: Summary of DM detection strategy (left), $\chi$ is DM, Q is the transfer momentum and q is the quark, and the typical collider signature of particles recoiled against a large MET (right).

## DARK MATTER COLLIDER SEARCHES

Usually the collider searches focus on leading generic Feynman Diagram responsible for Dark Matter production, specifically a pair production of WIMP pair plus the initial or final state radiation (ISR/FSR) of a gluon, photon or a weak gauge boson Z, W±. Different channels have different coupling to the WIMP. On the detector level, the ISR/FSR particle is needed to balance the two WIMP's momentum allowing the event to be "triggerable" as shown in Figure 3. Generally the selected event requires a central leading parton with high transverse momentum PT and a WIMP pair with high MET to forms a back-to-back topology. Additionally the angular separation Δφ between the leading parton and MET has to be more than 0.5 radians to minimize the contribution from fake missing energy from jet mismeasurements [11]. The common backgrounds to the analyses is the irreducible background Z(νν) + jets where the neutrinos decayed from the Z boson escape the detector undetected, and the W + jet where the decay lepton is mis-identified and therefore out of detector acceptance and therefore "lost". Nevertheless their contribution to the signal region can be estimated from the selected control samples from the signal event with a specific transfer function to account for kinematics differences [12].

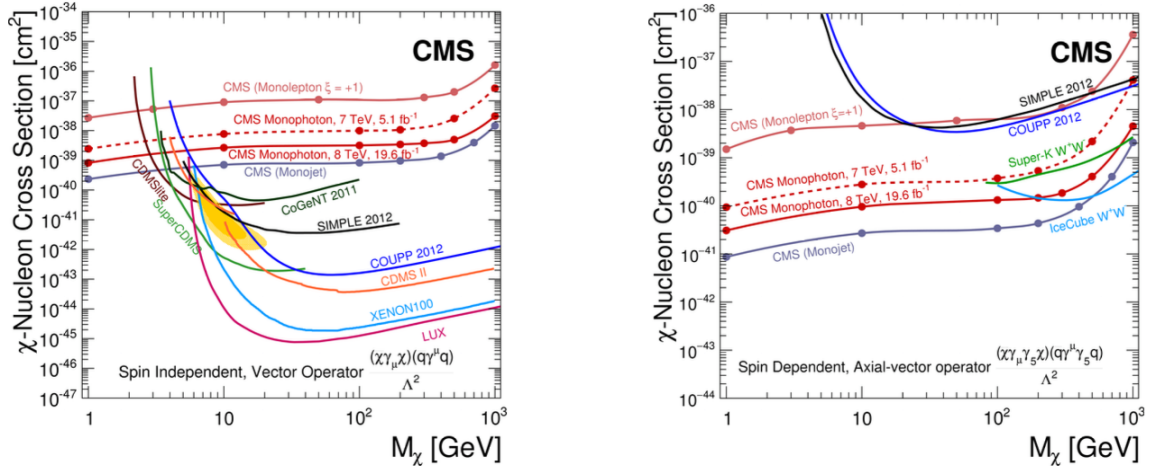

**FIGURE 4:** 90% CL upper limits on the DM-nucleon scattering cross section, from the monojet, monophoton and monolepton ($\xi = +1$) searches with vector operator, as a function of the dark matter mass, for spin independent (left) and spin-dependent (right) interactions. [13]

So far the CMS experiment's analyses revealed the collected data from the 8 TeV collisions in Run-1 at 19.5 pb−1 was similar with the background expectation, therefore a 90% Confident Level limits were derived for the three channels on the DM-nucleon scattering cross section [13]. In Figure 4 the collider limits were compared with the direct detection limits in spin-independent and spin dependent scenarios. The three channels were persistently searched into the low mass DM region where the direct detection fail due to the small recoil signals at low mass DM. On the other hand the collider analyses were complementing the spin-dependent interaction and excellently excluded most of the phase space. However at higher DM mass the LHC is incapable to produce such heavy DM, and therefore the production cross section dropped, so is the limits seen to be poor in exclusion.

The Monojet channel is expected to give the strongest contribution due to the rate of gluon and quark ISR is larger relative to other SM radiation as shown in Figure 5. The gluon and quarks are not free particles and further hadronized forming a spray of particle called jet which introduces higher uncertainties at higher MET.

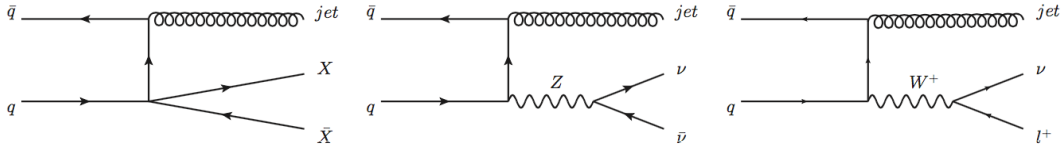

**FIGURE 5:** The Feynman Diagram of DM pair production in association with gluon (left), and the background $Z(\nu\nu)$ + jet (middle) and $W(l\nu)$ + jet (right).

Since the indirect detection is looking for an energetic gamma-ray, the Monophoton channel is a potential channel which consists of a final state of large MET and an energetic photon. The photon required to pass through a quality tight cut to ensure it is not an electron faking photon [12]. In terms of backgrounds, the Z + γ, W + γ and γ + multijets are the background processes complicating this channel. Despite the Monophoton does not couple strongly compared to Monojet, the fact that photon only interacts electromagnetically has given the advantage of being able to precisely measure from the electromagnetic calorimetry, thus reducing the experimental uncertainties.

The Mono vector boson channel is looking for final state consist an energetic vector gauge boson with a high MET. For example, the Mono-W signature is characterized by a single lepton with high MET whereas Mono-Z is reconstructed with a pair of charged lepton with high MET. The strength of gauge boson radiates off a quark pair initial state depends on parameter ξ which parametrized the relative strength of the coupling to down-quarks with respect to up-quarks [14]. Compared to Monojet the Mono-lepton channel is cleaner with small experimental systematic error and is likely to out-perform Monojet searches with increased luminosity or pile-up. On the other hand, the Mono-W searches also provide a valuable input on how to disentangle WIMP couplings to up-type versus down-type quarks.

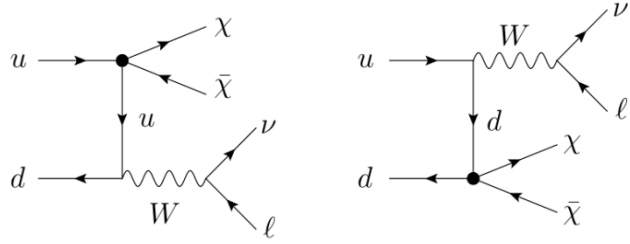

**FIGURE 6:** The interference of two scenarios where the W boson radiates from the u-quark or the d-quark or the d-quark [14].

## OUTLOOK

The compelling motivation for Dark Matter from the cosmological observation has led to the conclusion that new species of particles exist which is not described by the Standard Model. Thanks to the WIMP Miracle, the collider searches undertaken by the LHC have hailed as an ideal machine to produce DM particle by exploring promising signature such as Monojet, Monophoton and Mono-V. The LHC searches are well known in probing low DM mass and demonstrating strong interaction with axial vector DM in spin-dependent interaction scenario. During the 8TeV DM analysis, the EFT approach is commonly used due to the striking feature of being model-independent. It provide an economical way to learn about the underlying physics of a DM interaction without undermining the generic DM physics by forcing one into a model dependent fashion. However the EFT applicability is questionable which has discussed in section 3. Despite there exists a benchmark [10] to evaluate the validity of EFT description, the utilized unitarity technique is weaker [12]. According to the report proposed by ATLAS and CMS [15], the simplified model will be the focus of 13TeV DM analysis as it offers a complete theory of DM interaction. The EFT approach will still be essential because of the model independent interpretation of collider bound which is very much needed considering our current lack of knowledge towards the nature of DM particle and its interactions. In summary, the complementary collider searches harmonizing all the DM searches from Cosmology can therefore be considered as a concordance approach to improve the discovery potential.

## ACKNOWLEDGEMENT

The authors would like to acknowledge the financial support from the National Centre for Particle Physics (NCPP), and the University of Malaya Research Grant project number RP012B-13AFR (Detector Development And Electronics For Particle Physics). The authors would also wish to thank Dr. J. K. Komaragiri and Prof. Dr. W. A. T. Wan Abdullah from University of Malaya for useful discussion and supervision, and the CMS publishing committee for approval of the CMS results included in this paper.